\RequirePackage{fix-cm}
\documentclass[smallcondensed,a4paper]{svjour3}  

\smartqed  
\usepackage{graphicx}
\begin{document}

\title{Hill plot focusing on Ce compounds with high magnetic-ordering-temperatures and consequent study of Ce$_{2}$AuP$_{3}$ 
}

\titlerunning{Hill plot and consequent study of Ce$_{2}$AuP$_{3}$}        

\author{Junya Miyahara         \and
        Naoki Shirakawa       \and
        Yuta Setoguchi     \and
        Masami Tsubota    \and
        Kento Kuroiwa     \and
        Jiro Kitagawa     
}


\institute{Junya Miyahara  \at
              Department of Electrical Engineering, Faculty of Engineering, Fukuoka Institute of Technology, 3-30-1 Wajiro-higashi, Higashi-ku, Fukuoka 811-0295, Japan \\
           \and
           Naoki Shirakawa \at
             Flexible Electronics Research Center, National Institute of Advanced Industrial Science and Technology, Tsukuba, Ibaraki 305-8565, Japan \\
           \and
           Yuta Setoguchi \at
             Department of Electrical Engineering, Faculty of Engineering, Fukuoka Institute of Technology, 3-30-1 Wajiro-higashi, Higashi-ku, Fukuoka 811-0295, Japan \\
           \and
           Masami Tsubota \at
             Physonit Inc., 6-10 Minami-Horikawa, Kaita Aki, Hiroshima 736-0044, Japan \\
           \and
           Kento Kuroiwa \at
             Department of Electrical Engineering, Faculty of Engineering, Fukuoka Institute of Technology, 3-30-1 Wajiro-higashi, Higashi-ku, Fukuoka 811-0295, Japan \\
\and
           Jiro Kitagawa \at
             Department of Electrical Engineering, Faculty of Engineering, Fukuoka Institute of Technology, 3-30-1 Wajiro-higashi, Higashi-ku, Fukuoka 811-0295, Japan \\
           \email{j-kitagawa@fit.ac.jp}
}

\date{Received: date / Accepted: date}

\maketitle

\begin{abstract}
Hill plot is a well-known criterion of the $f$-electron element interatomic threshold-distance separating the nonmagnetic state from the magnetic one in actinides or lanthanides. We have reinvestigated the Hill plot of Ce compounds using a commercial crystallographic database CRYSTMET, focusing on a relationship between the Ce-Ce distance and the magnetic ordering temperature, because a Ce compound with no other magnetic elements scarcely has a magnetic ordering temperature higher than 20 K. The Hill plot of approximately 730 compounds has revealed that a Ce compound, especially for ferromagnet, showing the high magnetic-ordering-temperature would require a short Ce-Ce distance with a suppression of valence instability of Ce ion. Through the study, we had interest in Ce$_{2}$AuP$_{3}$ with the Curie temperature of 31 K. The ferromagnetic nature has been examined by a doping effect, which suggests a possible increase of magnetic anisotropy energy. 
\keywords{cerium compound \and Hill plot \and ferromagnetism}
\end{abstract}

\section{Introduction}
\label{intro}
Hill plot is the widely accepted criterion, proposing the prediction of magnetism and superconductivity in actinides and lanthanides.\cite{Hill:1970,Smith:PhysicaB1980}
Koelling reported the Hill plot of approximately 35 Ce-compounds, in the form of superconducting or magnetic ordering temperatures vs Ce interatomic distances, to explain the existence of a threshold distance separating nonmagnetic ground states from magnetic ones\cite{Koelling:PhysicaB1985}.
The Hill plot stimulates us from crystallographic aspects of Ce compounds showing magnetic orderings, because a Ce compound with no other magnetic elements scarcely has a high magnetic-ordering-temperature.
CeRh$_{3}$B$_{2}$ is a well-known ferromagnet with very high Curie-temperature $T_{C}$ of 115 K\cite{Dhar:JPC1981}.
Since the discovery of CeRh$_{3}$B$_{2}$, there have been intensive researches on Ce compound with a high magnetic-ordering-temperature.

It would be important to shed light on a guideline of material design for such a Ce-compound.
Because the Ce-compounds investigated by Koelling are restricted to pure metal and binary compounds, we have reinvestigated Hill plot, surveying also ternary compounds.
Employing CRYSTMET: a commercial database of the structural and powder patterns of metals and intermetallics, a correlation between the shortest Ce-Ce distance and the magnetic ordering temperature has been investigated for approximately 730 compounds.
As understood by the reported Hill plot, Ce compounds tend to show intermediate valence (IV) states with decreasing Ce-Ce distance.
Nonetheless, the region of Ce-Ce distance for compounds showing high magnetic-ordering-temperatures above 20 K overlaps that of IV compounds.
The Hill plot based on CRYSTMET would propose a rather short Ce-Ce distance with the suppression of valence instability as the guideline of obtaining a Ce compound, especially of ferromagnetic type, showing a high magnetic-ordering-temperature. 

Through the crystallographic consideration of many Ce-compounds, we focused on Ce$_{2}$AuP$_{3}$ with $T_{C}$ of 31 K\cite{Eschen:ZAAC2001}.
Although Ce$_{2}$AuP$_{3}$ is metallic, the La counterpart shows a semiconducting behavior\cite{Eschen:ZAAC2001}.
This suggests that the density of states near the Fermi level of Ce$_{2}$AuP$_{3}$ may be sensitive to a doping.
The magnetic anisotropy energy of a ferromagnet can be tuned by a doping effect\cite{Enkovaara:PRB2003,Larson:PRB2004,Sakuma:JPSJ2013}, mainly due to the variation of density of states near the Fermi level.
Then we expect an easy tuning of magnetic anisotropy energy with doping in Ce$_{2}$AuP$_{3}$.
Furthermore, the doping may shift $T_{C}$ with varying Ce-Ce distance.

In the literature\cite{Eschen:ZAAC2001}, Ce$_{2}$AuP$_{3}$ was prepared by a direct reaction of constituent elements in an evacuated quartz tube.
Because a direct reaction in this study resulted in a broken quartz tube or production of CeP for some unknown reason, we have tried another preparation method.
There are many studies, reporting the crystal growth of phosphides by employing flux methods\cite{Jeitschko:ACTA1977,Nientiedt:JSSC1999,Schellenberg:ZNAT2010,Pfannenschmidt:MFC2011}. 
We have also tried the sample preparation by Sn, Pb and NaCl/KCl flux methods, in which the NaCl/KCl flux method leads to the reproducible synthesis of Ce$_{2}$AuP$_{3}$.
Following the synthesis of parent compound, we have prepared several doped-compounds and characterized the magnetic properties.
This paper discusses the crystallographic features of Ce compounds showing magnetic orderings based on the Hill plot, and reports the synthesis and the investigation of magnetic properties of the parent and doped samples.

\section{Materials and Methods}
We employed CRYSTMET (Toth Information Systems, ver. 5.7.0) to collect the crystallographic data of 724 Ce-compounds.
The surveyed compounds with the ID codes in CRYSTMET are summarized in Table S1 of the Supplementary Information.
The magnetic ordering temperature and the ordered type are also shown in Table S1.

Filings of Ce (99.9\%), Au powder (99.9\%) and P powder (99.999\%) with the stoichiometric composition were homogeneously mixed together in a glove box filled with argon gas.
The mixed powder ($\sim$0.5 g) was added to NaCl/KCl powder (the molar ratio 1:1, $\sim$2 g).
The mixture was placed in a carbon crucible in an evacuated quartz tube, which was heated to 950$^{\circ}$C with the rate 50$^{\circ}$C/h, held at that temperature for 48h, and then furnace cooled.
The product was washed by water to remove the NaCl/KCl flux, and obtained in a powder form.
The doped samples, Ce$_{2}$AuP$_{2.7}$Si$_{0.3}$, Ce$_{2}$AuP$_{2.7}$S$_{0.3}$ and Ce$_{2}$Au$_{0.9}$Pt$_{0.1}$P$_{3}$ were prepared, following the same manner.
The Si (99.999\%), S (99.99\%) and Pt (99.9\%) powders were used.
The samples were evaluated using a powder X-ray diffractometer (Shimadzu, XRD-7000L) with Cu-K$\alpha$ radiation. 

The temperature dependence of dc magnetic susceptibility $\chi_{dc}$(T) between 5 K and 50 K and the magnetization curve were checked by a Quantum Design MPMS. 
The temperature dependence of ac magnetic susceptibility $\chi_{ac}$(T) in an alternating field of 5 Oe at 100 Hz, between 0.48 K and 35 K, was also measured using the MPMS equipped with iHelium3\cite{Shirakawa:Poly2005,Shirakawa:JPCS2009}.Another measurement of $\chi_{ac}$(T) in an alternating filed of 8 Oe at 800 Hz, between 4 K and 300 K, was carried out using a closed-cycle He gas cryostat.

\label{f1}
\begin{figure}
\begin{center}
\includegraphics[width=9.5cm]{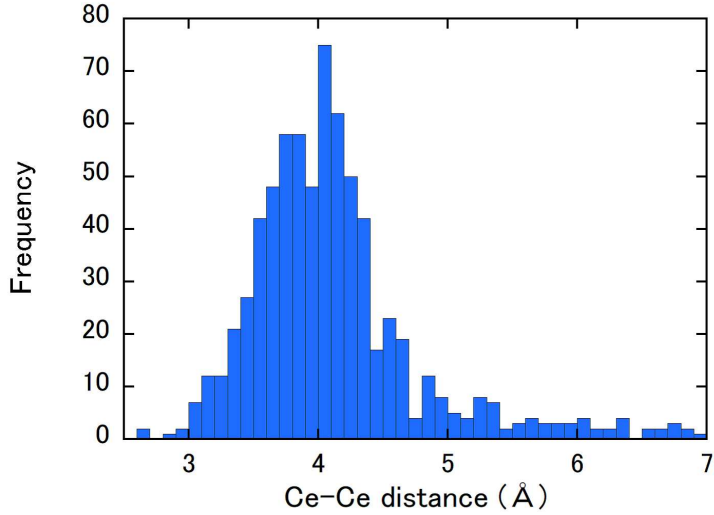}
\end{center}
\caption{Histogram of Ce-Ce distance for 724 Ce-compounds.}
\end{figure}

\section{Results and discussion}
Figure 1 shows the histogram of the shortest Ce-Ce distance for our surveyed compounds including both magnetic and nonmagnetic ground states.
The average distance is approximately 4.161 \AA.
The increase of Ce-Ce distance above the average one decreases the frequency steeply.
The frequency slightly decreases with decreasing Ce-Ce distance from 4.161 \AA \hspace{1.5mm} to 3.5 \AA, below which the sudden decrease of frequency is observed.
The Hill plots are shown in Figs.\ 2(a) and 2(b) for ferromagnets and antiferromagnets, respectively, with the vertical broken-line indicating the average Ce-Ce distance.
The horizontal solid line in each figure is drawn at $T_{C}$ (N\'eel temperature $T_{N}$) of 20 K.
To check that the valence instability tends to occur at shorter Ce-Ce distance, the fraction of IV compounds is calculated as shown in Fig.\ 3.
We eliminate the fractions below 3 \AA \hspace{1.5mm} and above 4.7 \AA, where the frequencies of compounds are small. 
One can see the rough increase of fraction of IV compounds by decreasing the Ce-Ce distance below the average one, which would be due to the general trend of increased hybridization between 4$f$ electrons and ligand ones.
Nonetheless, the region of Ce-Ce distance for compounds with high magnetic-ordering-temperatures above 20 K overlaps that of IV compounds. 
The tendency is noteworthy for ferromagnets. 
It is also noted that the increase of Ce-Ce distance above 4.6 \AA \hspace{1.5mm} depresses magnetic ordering temperature.
The Hill plot based on CRYSTMET would propose a rather short Ce-Ce distance with the suppression of valence instability as the guideline of obtaining a Ce compound, especially of ferromagnet, showing a high magnetic-ordering-temperature above 20 K.

\label{f2}
\begin{figure}
\begin{center}
\includegraphics[width=12cm]{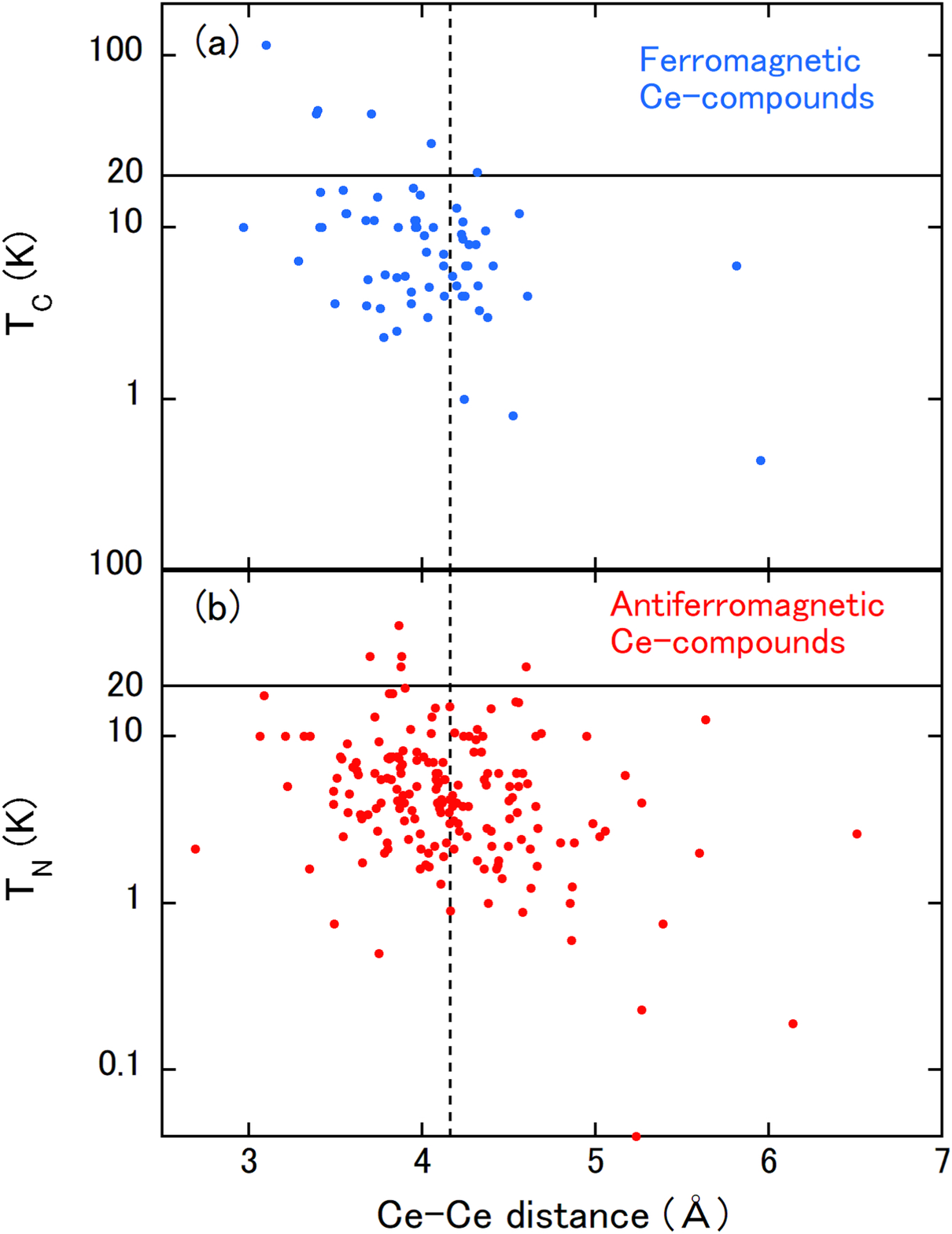}
\end{center}
\caption{Hill plots of Ce compounds showing magnetic orderings for (a) ferromagnets and (b) antiferromagnets.}
\end{figure}

\label{f3}
\begin{figure}
\begin{center}
\includegraphics[width=8.5cm]{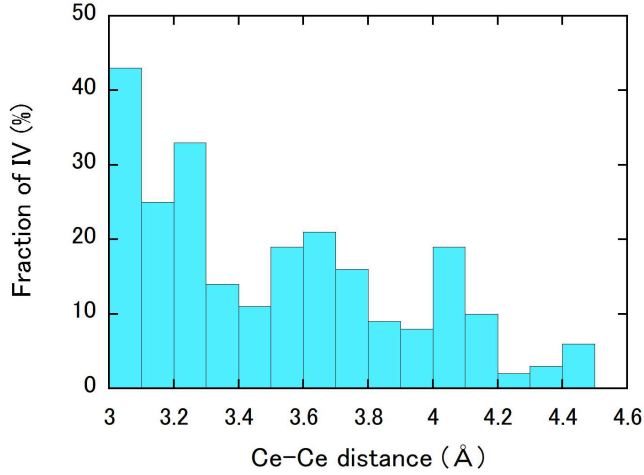}
\end{center}
\caption{Histogram of fraction of IV compounds.}
\end{figure}

\begin{table}[t]
\caption{Comparison of crystallographic and magnetic properties among Ce compounds with high magnetic-ordering-temperatures above 18 K. FM and AFM mean ferromagnetic and antiferromagnetic orderings, respectively.}
\label{t1}
\small
\begin{tabular}{cccccc}
\hline
Compound & Crystal system & Ce-Ce  & Magnetic & Ordering & Ref. \\
&& distance (\AA) & ordering type & temperature (K) & \\
\hline
CeRh$_{3}$B$_{2}$ & hexagonal & 3.098 & FM & 115 & \cite{Dhar:JPC1981}  \\
Ce$_{5}$NiPb$_{3}$ & hexagonal & 3.396 & FM & 48 & \cite{Goruganti:JAP2009}  \\
Ce$_{5}$Pb$_{3}$O & tetragonal & 3.706 & FM & 46 & \cite{Macaluso:CM2004}  \\
Ce$_{5}$CuPb$_{3}$ & hexagonal & 3.388 & FM & 46 & \cite{Tran:JSSC2007} \\
CeScGe & tetragonal & 3.864 & AFM & 46 & \cite{Singh:JPCM2001} \\
Ce$_{2}$AuP$_{3}$ & orthorhombic & 4.052 & FM & 31 & \cite{Eschen:ZAAC2001} \\
CeZn & cubic & 3.697 & AFM & 30 & \cite{Kadomatsu:PRB1986} \\
CeC$_{2}$ & tetragonal & 3.881 & AFM & 30 & \cite{Sakai:JCP1981} \\
CeBi& cubic & 4.600 & AFM & 26 & \cite{Nereson:JAP1971}  \\
CeScSi& tetragonal & 3.875 & AFM & 26 & \cite{Singh:JPCM2001}  \\
CeCoGe$_{3}$& tetragonal & 4.316 & FM & 21 & \cite{Pecharsky:PRB1993}  \\
CeMg& cubic & 3.901 & AFM & 19.5 & \cite{Pierre:JMMM1984}  \\
CeNiC$_{2}$& orthorhombic & 3.831 & AFM & 18 & \cite{Pecharsky:PRB1998}  \\
CeCoC$_{2}$& monoclinic & 3.810 & AFM & 18 & \cite{Pecharsky:PRB1998}  \\
\hline
\end{tabular}
\end{table}

Table 1 shows the comparison of crystallographic and magnetic properties among Ce compounds with magnetic ordering temperatures higher than 18 K in our surveyed data.
We note here that most compounds in the table possess high crystal-system symmetry such as cubic, tetragonal and hexagonal.
We have focused on Ce$_{2}$AuP$_{3}$, crystallizing into an orthorhombic U$_{2}$NiC$_{3}$-type structure with $T_{C}$ of 31 K\cite{Eschen:ZAAC2001}.
As mentioned in the introduction, a possible doping-sensitive density of states in Ce$_{2}$AuP$_{3}$ might offer an easy tuning of magnetic anisotropy energy. 
Another interesting point is that the crystal system symmetry of Ce$_{2}$AuP$_{3}$ is the lowest among the compounds with magnetic ordering temperatures higher than 30 K.
The magnetic anisotropy energy is generally related with the crystal symmetry.
A lower crystal symmetry tends to lead to more tunable lattice parameters upon doping, which also motivated us to examine a doping effect on Ce$_{2}$AuP$_{3}$.
In addition, a possible shift of $T_{C}$ might be expected, if the Ce-Ce distance can be largely altered.
Before starting the study, we have tried a synthetic route of parent compound different from that described in the literature\cite{Eschen:ZAAC2001} and have found that the NaCl/KCl flux method leads to the reproducible production of sample.

\label{f4}
\begin{figure}
\begin{center}
\includegraphics[width=9.5cm]{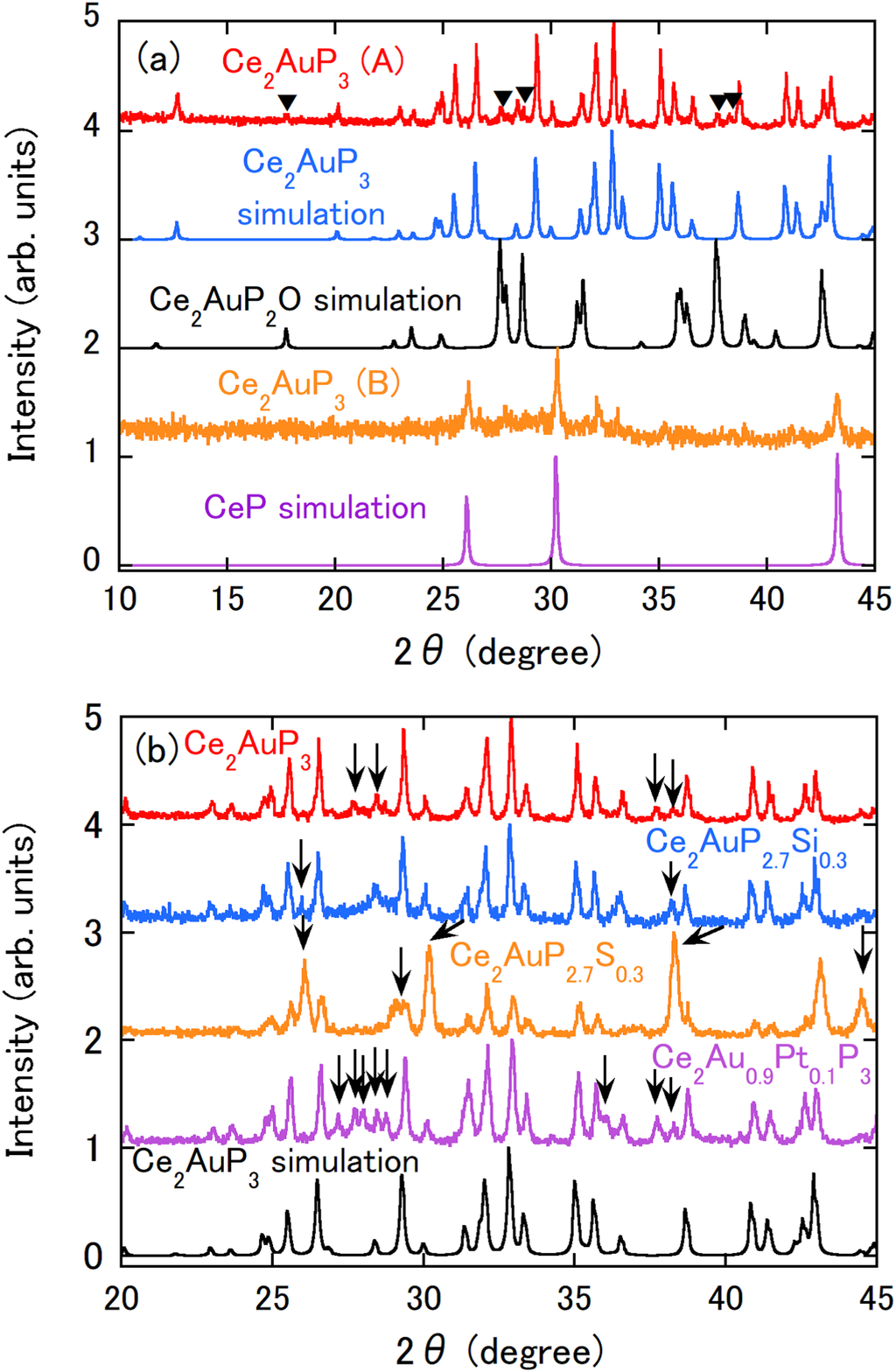}
\end{center}
\caption{(a) XRD patterns of Ce$_{2}$AuP$_{3}$ prepared with (A) flux and (B) without flux. The simulated XRD patterns of Ce$_{2}$AuP$_{3}$, Ce$_{2}$AuP$_{2}$O and CeP are also shown. The origin of each pattern has been shifted by an integer value for clarity. (b) XRD patterns of parent and doped Ce$_{2}$AuP$_{3}$. The simulated XRD pattern of Ce$_{2}$AuP$_{3}$ is also shown. The origin of each pattern has been shifted by an integer value for clarity.}
\end{figure}

Figure 4(a) shows the X-ray diffraction (XRD) pattern of Ce$_{2}$AuP$_{3}$ (A) prepared with flux, which is compared with the simulated pattern.
Although they agree with each other, small amount of impurity phases (triangles in Fig.\ 4(a)) mainly ascribed to Ce$_{2}$AuP$_{2}$O appear. 
Ce$_{2}$AuP$_{2}$O is an antiferromagnet with $T_{N}$ of 13.1 K\cite{Bartsch:IC2013}.
In order to check the need of NaCl/KCl flux, we synthesized Ce$_{2}$AuP$_{3}$ by the same heat treatment but without a flux. 
Figure 4(a) also displays the XRD pattern of obtained sample (Ce$_{2}$AuP$_{3}$ (B)), which is almost identical to that of CeP. 
Therefore, the NaCl/KCl flux seems to enhance the crystal growth of Ce$_{2}$AuP$_{3}$ greatly.

The XRD patterns of doped samples Ce$_{2}$AuP$_{2.7}$Si$_{0.3}$, Ce$_{2}$AuP$_{2.7}$S$_{0.3}$ and Ce$_{2}$Au$_{0.9}$Pt$_{0.1}$P$_{3}$ are shown in Fig.\ 4(b).
In the Si doping, the appearance of another impurity phases is signaled by arrows in Fig.\ 4(b).  
For the S-doped sample, impurity phases, partially assigned as CeP, denoted by arrows dominate over the parent one. 
Ce$_{2}$Au$_{0.9}$Pt$_{0.1}$P$_{3}$ shows many impurity peaks.
The lattice parameters of prepared samples were refined by the Rietveld refinement program RIETAN FP\cite{Izumi:SSP2007}, adding a recently proposed criterion for obtaining accurate lattice parameters\cite{Tsubota:SR2017}, as listed in Table 2.
The lattice parameters would be slightly increased by the Si (S) doping.
The partial replacement of Au by Pt contracts the lattice parameters.
The XRD results with the increased amount of impurity phases indicate an incomplete replacement of P by Si (S) or Au by Pt atoms.\  

\begin{table}[t]
\caption{Lattice parameters of prepared samples.}
\label{t2}
\small
\begin{tabular}{cccc}
\hline
Compound & $a$ (\AA) & $b$ (\AA) & $c$ (\AA) \\
\hline
Ce$_{2}$AuP$_{3}$& 7.7419(5) & 4.2109(2) & 16.1237(10) \\
Ce$_{2}$AuP$_{2.7}$Si$_{0.3}$& 7.7510(7) & 4.2115(4) & 16.1315(15) \\
Ce$_{2}$AuP$_{2.7}$S$_{0.3}$& 7.7463(13) & 4.2118(7) & 16.1328(24) \\
Ce$_{2}$Au$_{0.9}$Pt$_{0.1}$P$_{3}$& 7.7368(6) & 4.2072(3) & 16.1051(13) \\
\hline
\end{tabular}
\end{table}

\label{f5}
\begin{figure}
\begin{center}
\includegraphics[width=9.5cm]{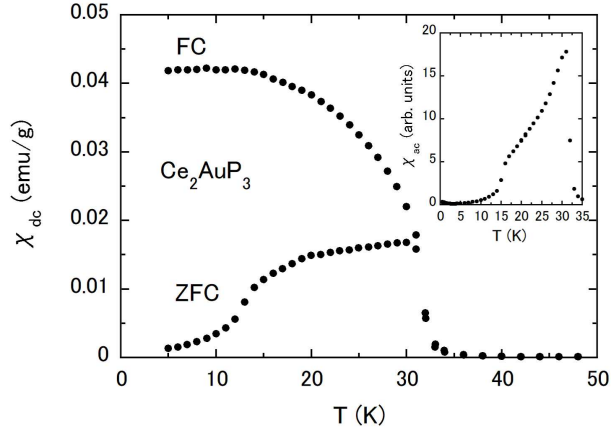}
\end{center}
\caption{Temperature dependence of $\chi_{dc}$ of Ce$_{2}$AuP$_{3}$. The inset is the temperature dependence of $\chi_{ac}$ of Ce$_{2}$AuP$_{3}$.}
\end{figure}

\label{f6}
\begin{figure}
\begin{center}
\includegraphics[width=8.5cm]{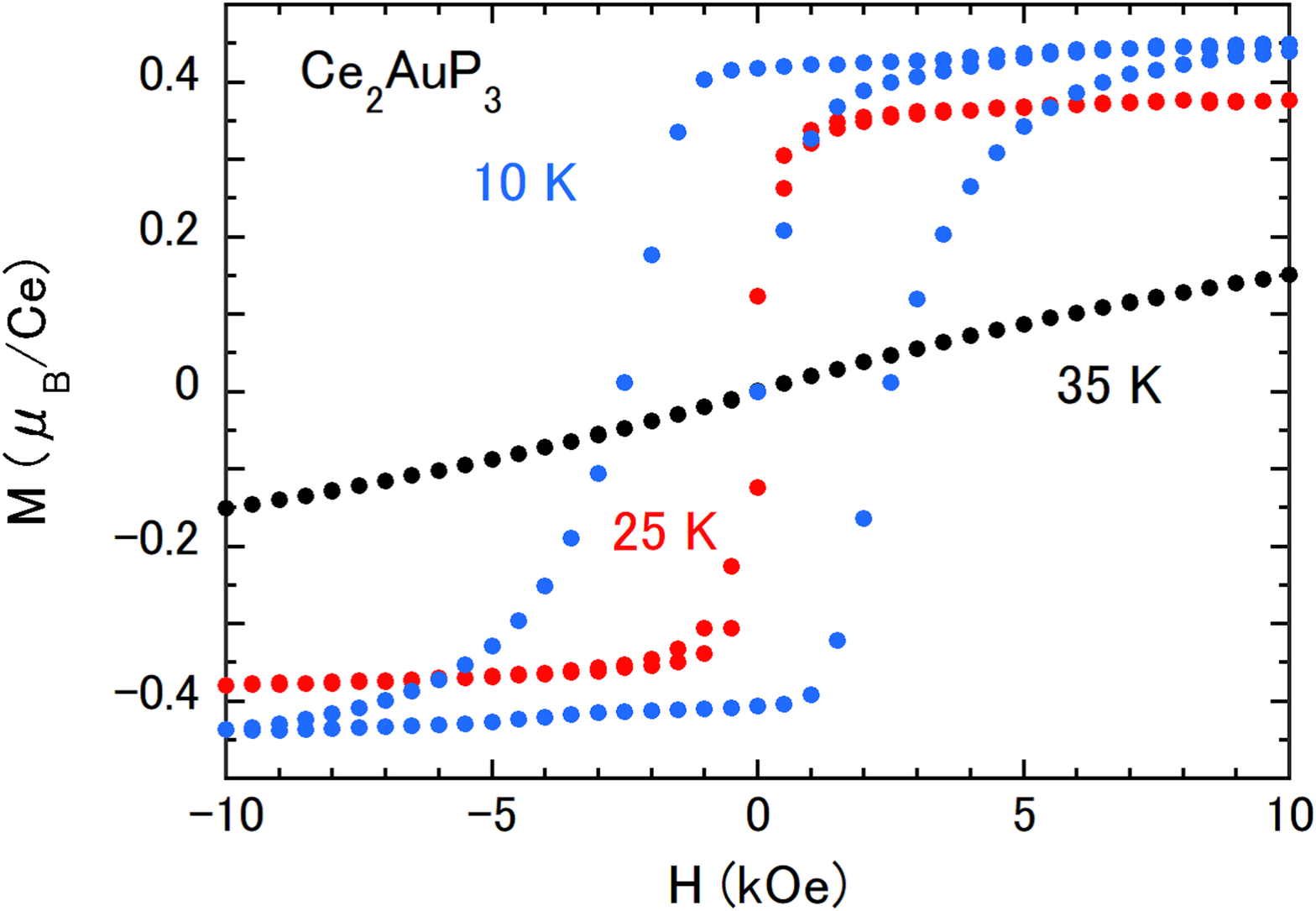}
\end{center}
\caption{Magnetization curves of Ce$_{2}$AuP$_{3}$.}
\end{figure}

Figure 5 shows $\chi_{dc}$ (T) of the parent compound measured after cooling sample in zero magnetic field (ZFC) and by cooling the sample in the magnetic field (FC).
The applied magnetic field was 20 Oe.
$\chi_{dc}$ in FC shows a typical ferromagnetic behavior and well agrees with that reported in the literature\cite{Eschen:ZAAC2001}.
$T_{C}$ determined by the minimum of $d\chi_{dc}/dT$ is approximately 31 K.
The inset of Fig.\ 5 exhibits $\chi_{ac}$ (T) of the parent compound with a sharp peak at approximately 31 K.
At approximately 15 K, $\chi_{ac}$ (T) shows a shoulder-like anomaly, which suggests a spin reorientation as in the case of DyAl$_{2}$\cite{Lima:PRB2005}.
As shown in Fig.\ 6, the magnetization $M$ of Ce$_{2}$AuP$_{3}$ at 35 K shows a paramagnetic behavior up to 10 kOe. 
Below $T_{C}$, $M$ shows a hysteresis loop, growing in size with decreasing temperature.

$\chi_{ac}$ (T) of doped samples are summarized in Figure 7 with that of the parent compound, demonstrating no shift of $T_{C}$.
The small peak at approximately 13 K observed in Ce$_{2}$Au$_{0.9}$Pt$_{0.1}$P$_{3}$ reflects the presence of Ce$_{2}$AuP$_{2}$O phase as evidenced in Fig.\ 4(b).
$\chi_{ac}$ (T) of Ce$_{2}$AuP$_{2.7}$S$_{0.3}$ is almost identical to that of the parent compound.
On the other hand, in Ce$_{2}$AuP$_{2.7}$Si$_{0.3}$ or Ce$_{2}$Au$_{0.9}$Pt$_{0.1}$P$_{3}$, the shoulder-like anomaly associated with the spin reorientation is depressed and the peak at $T_{C}$ becomes sharper compared to that of the parent compound.
$\chi_{ac}$ is correlated with the reversible initial magnetization process.
The disappearance of spin reorientation phenomenon and the sharper $\chi_{ac}$-peak at $T_{C}$ support the increased difficulty of reversible process, which means an increased magnetic anisotropy energy.
Considering that the Si- (Pt-) and S-dopings correspond to the hole and electron dopings, respectively, only the hole doping would affect the magnetic anisotropy energy.

\label{f7}
\begin{figure}
\begin{center}
\includegraphics[width=9.5cm]{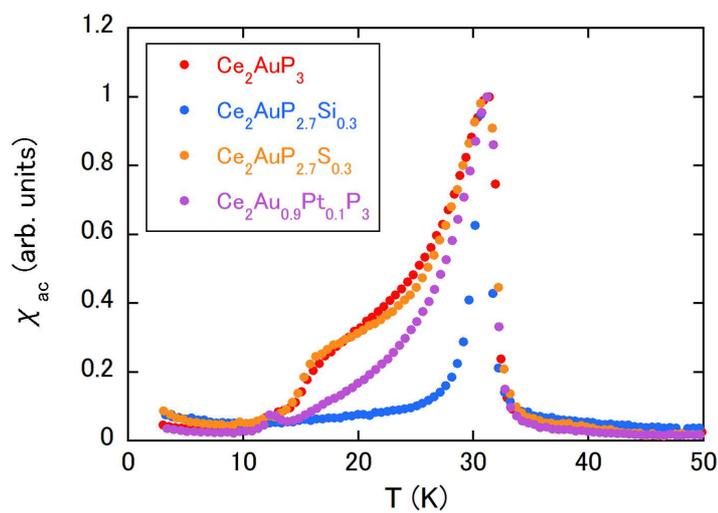}
\end{center}
\caption{Temperature dependences of $\chi_{ac}$ of parent and doped Ce$_{2}$AuP$_{3}$.}
\end{figure}

\section{Summary}
We have reinvestigated the Hill plot focusing on the correlation between the magnetic ordering temperature and the Ce-Ce distance.
Surveying 724 Ce-compounds based on CRYSTMET data, we have proposed that a Ce-compound with a high magnetic-ordering-temperature would require a short Ce-Ce distance with the suppression of valence instability of Ce ion.
Through the study of Hill plot, we have focused on Ce$_{2}$AuP$_{3}$ with high $T_{C}$.
Due to the possible doping-sensitive density of states in Ce$_{2}$AuP$_{3}$, we have expected the easy tuning of magnetic anisotropy energy by doping effect.
We have found that the sample can be reproducibly obtained by the NaCl/KCl flux method.
While the amount of impurity phases increases in each doped-sample, the lattice parameters would be slightly changed by the doping effect.
We have found that the spin reorientation would occur at approximately 15 K in the parent compound.
While $\chi_{ac}$ (T) of the S-doped sample is almost identical to that of the parent one, $\chi_{ac}$ (T) of Ce$_{2}$AuP$_{2.7}$Si$_{0.3}$ or Ce$_{2}$Au$_{0.9}$Pt$_{0.1}$P$_{3}$ shows the depression of the spin reorientation, and the sharper peak at $T_{C}$ compared to that of the parent compound.
These results imply the possible increased magnetic anisotropy energy by the hole doping.


%
%

\begin{acknowledgements}
This work was supported by the Comprehensive Research Organization of Fukuoka Institute of Technology.
\end{acknowledgements}



\end{document}